# 基于路径语义表示的静态警报自动确认方法

张俞炜 [1,2], 邢颖 [3], 李戈 [1,2], 金芝 [1,2]

[1](北京大学 计算机学院,北京 100871)
[2](高可信软件技术教育部重点实验室(北京大学),北京 100871)
[3](北京邮电大学 人工智能学院,北京 100876)
通讯作者: 金芝, E-mail: zhijin@pku.edu.cn

**摘　要**: 尽管静态分析工具能够在软件开发生命周期的早期阶段帮助开发人员检测软件中的潜在缺陷,但该类工具往往存在警报假阳性率高的问题.为了提高该类工具的可用性,研究人员提出许多警报确认技术来对假阳性警报进行自动分类.然而,已有方法大多利用手工设计的特征或语句级的抽象语法树标记序列来表示缺陷代码,难以从报告的警报中捕获语义.为了克服传统方法的局限性,本文利用深度神经网络强大的特征抽取和表示能力从控制流图路径中学习代码语义表征用于警报确认.控制流图是程序的执行过程抽象表示,因此控制流图路径序列能够引导模型更精确地学习与潜在缺陷相关的语义信息.本文通过微调预训练语言模型对路径序列进行编码并从中捕获语义特征用于模型构建,最后在 8 个开源项目上与最先进的基线方法进行大量对比实验验证了所提方法的有效性.
**关键词**: 警报自动确认;路径分析;深度学习;预训练语言模型
中图法分类号: TP311

## Automated Static Warning Identification via Path-based Semantic Representation

ZHANG Yu-Wei[1,2], XING Ying[3], LI Ge[1,2], JIN Zhi[1,2]

[1](School of Computer Science , Peking University, Beijing 100871, China)
[2](Key Laboratory of High Confidence Software Technologies (Peking University), Ministry of Education, Beijing 100871, China)
[3](School of Artificial Intelligence, Beijing University of Posts and Telecommunications, Beijing 100876, China)

**Abstract**: Despite their ability to aid developers in detecting potential defects early in the software development life cycle, static analysis tools often suffer from precision issues (i.e., high false positive rates of reported alarms). To improve the availability of these tools, many automated warning identification techniques have been proposed to assist developers in classifying false positive alarms. However, existing approaches mainly focus on using hand-engineered features or statement-level abstract syntax tree token sequences to represent the defective code, failing to capture semantics from the reported alarms. To overcome the limitations of traditional approaches, this paper employs deep neural networks' powerful feature extraction and representation abilities to generate code semantics from control flow graph paths for warning identification. The control flow graph abstractly represents the execution process of a given program. Thus, the generated path sequences of the control flow graph can guide the deep neural networks to learn semantic information about the potential defect more accurately. In this paper, we fine-tune the pre-trained language model to encode the path sequences and capture the semantic representations for model building. Finally, this paper conducts extensive experiments on eight open-source projects to verify the effectiveness of the proposed approach by comparing it with the state-of-the-art baselines.

　　软件是推动新一代信息技术发展的驱动力,以大型软件系统为核心的计算机应用逐渐在国防安全、航空航天以及金融等领域中发挥着越来越重要的作用.但是,随着软件规模和复杂性的急剧膨胀,开发人员也面临着保证软件高质量的挑战.目前,开发人员通常依赖自动化静态分析工具[1-3](ASAT, automatic static analysis tool)来扫描代码库,并在软件开发的早期阶段发现潜在的缺陷(defect).ASAT 能够在不运行程序的情况下通过匹配代码和先验缺陷模式知识来检测代码中可能的属性违反.然而,ASAT 在面对大规模软件系统时往往需要在精度和性能之间进行权衡.因此,该类工具并不能保证所报告的警报(warning)都是真实缺陷.已有工作[4-7]表明 ASAT 的警报假阳率可能达到 30%~100%,许多开发人员也认为 ASAT 所报告的警报与缺陷并不相关[8-10].目前,工业界大多采用人工查看的方法对 ASAT 报告的警报进行逐个确认,而开发人员只会修复被确认为是真实缺陷的警报[11].静态警报确认不仅需要开发人员具备专业级知识,人工确认的难度也会随着 ASAT 报告的警报数量的

上升而增加[12-14].因此,研究静态警报自动确认技术以帮助开发人员保证软件质量是非常重要的.

近年来,研究人员利用机器学习来自动学习警报的假阳性警报模式,有效地降低了警报确认的人工成本.现有的基于机器学习的警报自动确认模型[12,15]通常需要首先设计基于代码复杂度、开发过程历史修改信息以及警报信息的静态度量[16]来构建相关的数据集.但人工设计的静态度量通常难以表示警报所对应程序的复杂语法结构信息及语义特征,从而导致警报自动确认模型精度缺失[17-19].此外,在新开发的软件项目中通常缺乏足够的历史数据来训练高精度的警报确认模型.为此,研究人员提出利用已有项目(源项目)的历史警报数据训练模型并在新的项目(目标项目)上进行预测,但源项目和目标项目之间往往存在数据分布差异[20].因此,提高在跨项目场景下警报确认模型的泛化能力也是亟待解决的关键问题.

随着人工智能领域中深度学习技术的兴起,研究人员也开始尝试在结构更为复杂的数据上构造深度学习模型来解决软件工程领域中的问题[21].已有研究[22-24]探索利用深度神经网络(DNN, deep neural network)从程序的抽象语法树(AST, abstract syntax tree)中自动学习丰富的语法结构信息.与传统的基于代码静态度量的研究相比,DNN强大的表示学习能力能够有效替代人工设计特征的过程.然而,导致潜在缺陷的原因往往隐藏在程序的语义中,并且只在特定条件下才会引发运行时异常输出[25].因此基于代码静态度量和AST标记(token)序列的特征表示可能还不足以揭示程序中的缺陷.如图1所示,两个ASAT所报告的C语言程序静态警报示例代码中都分别包含一个twoIntsStruct变量类型声明语句,一个if条件语句以及一个printLine函数调用语句.两段代码只在if条件语句的谓词表达式中使用了不同的运算符.由于函数bad在if条件语句中使用的按位与运算符(&)会导致谓词表达式两边都被执行,因此第4行中ASAT所报告的警报1会引发空指针解引用缺陷.而函数good通过在if条件语句中使用逻辑与运算符(&&)可以修复该缺陷,即第10行中的警报2是一个假阳性警报.若使用传统的静态代码度量作为特征表示,两段警报代码的特征向量是完全相同的,因为两段代码有相同的代码行、Halstead度量元等.同样地,现有的基于AST表示的方法[19,23]往往忽略了其叶子节点中的原子指令信息.在这个例子中,两段警报代码都将被表示为相同的语句级token向量[VariableDeclarator, IfStatement, MethodInvocation],并没有包含具体的运算符信息(即原子指令).因此,上述方法均无法判断出假阳性警报.

```
1   void bad()
2   {
3       twoIntsStruct *twoInts = NULL;
4       if ((twoInts != NULL) & (twoInts->intOne == 5)) //警报1
5           printLine("intOne == 5");
6   }
```

```
7   void good()
8   {
9       twoIntsStruct *twoInts = NULL;
10      if ((twoInts != NULL) && (twoInts->intOne == 5)) //警报2
11          printLine("intOne == 5");
12  }
```

图 1  警报代码示例

针对上述问题,本文提出一种融合控制流图(CFG, control flow graph)路径与预训练语言模型的静态警报自动确认方法 SWIPER(Static Warning Identification via Path-based sEmantic Representation).首先,将警报代码所在源文件解析为对应的CFG,并通过CFG生成从入口节点到警报确认点(IP, identification point)的可达执行路径.与AST相比,CFG能够表示程序的执行过程,且CFG内的每个节点都能与对应的AST节点相匹配,并包含原子指令.因此,基于CFG的代码表示可以有效地区分程序代码中细微的差异,即图1中两个不同的运算符(&和&&)在解析后能够转化为不同的原子指令.同时,已有研究[26-28]表明路径敏感性分析是检测缺陷时消除假阳性警报的有效技术.因此,从警报代码中所生成的CFG路径能够更精确地表示与潜在缺陷相关的语义信息.接着,通过利用基于频率相似性检索的实例选择方法来筛选源项目中与目标项目相关的警报实例用于后续模型训练,从而缓解由于不同项目之间的数据分布差异导致的负迁移问题.最后,本文使用针对编程语言的开源预训练模型 CodeBERT[29]从标记化的 CFG 路径中自动学习语义表示以训练静态警报自动确认模型.CodeBERT 是基于自然语言和编程语言的双模态预训练模型,可以捕获二者之间的语义联系并输出支持代码理解任务的通用表示.本文将提出的方法应用在特定的下游任务(即警报确认)上对CodeBERT进行微调(fine-tuning)来达到良好的模型泛化性能,从而缓解跨项目场景下警报确认模型精度低的问题.

本文第1节介绍静态警报确认的相关工作.第2节介绍本文所提出的静态警报自动确认方法 SWIPER.第3节介绍本文的实验设置.第4节通过展示对比实验结果验证了所提方法的有效性.第5节进行总结与展望.

## 1　相关工作

当前静态警报自动确认的研究热点在于挖掘警报代码中与潜在缺陷相关的语义信息以构建精确的警报确认模型,并通过缩小不同软件项目间的数据分布差异来提高跨项目场景下警报确认模型的性能.结合本文所研究的具体内容,本节将对静态警报确认领域中以下几个方面的研究工作进行回顾与分析.

### 1.1　基于机器学习的静态警报自动确认技术

自动化的静态警报自动确认技术是软件工程领域中非常重要的研究课题.基于机器学习方法[12,15]进行警报确认可以从软件项目的历史数据中学到静态代码度量和潜在缺陷之间的关系,从而有效地减轻开发人员对 ASAT 所报告的警报进行人工审查所需要的时间成本.Pan 等人[30]提出面向 C 语言特性的程序切片指标来衡量程序的复杂度,并利用贝叶斯网络分类器来计算 ASAT 所报告的警报是真实缺陷的概率.Ruthruff 等人[31]使用文件度量、代码搅动信息、警报信息等特征建立逻辑回归模型来预测由 FindBugs 所报告的警报的类别.此外,Ruthruff 等人还通过排除低预测性能的特征来提高模型的预测性能.Heckman 和 Williams[32]提出了一种假阳性警报消除模型的构建流程.针对不同的软件项目,首先从 51 个候选警报特征中选择特征子集.接着利用 15 种机器学习算法在所选特征子集上构建模型以对警报进行分类.Hanam 等人[33]首先利用每个警报 IP 处的上下文代码所对应的 AST 设计代码模式,然后根据所设计的模式构建特征向量作为机器学习分类器的输入,最后建立警报确认模型对新的警报进行分类.类似地,Yoon 等人[34]利用从 AST 中所提取的警报结构特征作为支持向量机分类器的输入来计算新报告的警报是真实缺陷的概率.Zhang 等人[35]通过对导致潜在缺陷的函数变量进行定值-引用分析,设计出一组细粒度的特征用于警报确认模型构建.实验结果表明,与传统的代码静态度量相比,所提出的变量级特征能够有效提高警报确认任务的模型性能.与上述基于机器学习的警报确认技术相比,本文所提出的 SWIPER 框架利用深度学习技术来自动生成源代码的语义特征,以代替传统机器学习中人工提取特征的过程.

### 1.2　基于深度学习的代码语义特征自动生成技术

近几年来,随着开源平台(比如 GitHub)的兴起,研究人员比以往更容易获取大规模的优质代码数据.这使得研究人员开始尝试利用大规模代码和深度学习模型来解决静态警报确认这一自动化任务.Koc 等人[18]应用循环神经网络(RNN, recursive neural network)和图神经网络(GNN, graph neural network)从 Java 字节码以及静态分析警报文本信息中自动学习语义特征表示,以构建分类器去滤假阳性警报.Lee 等人[36]训练基于卷积神经网络(CNN, convolutional neural network)的分类器以便从引发 ASAT 报告警报的程序代码中学习特殊词汇模式,从而对假阳性警报进行自动确认.Xia 等人[37]提出了一种基于日志挖掘的警报自动确认方法,该方法利用 DNN 来预测由 Doc2Vec 转化的日志序列特征向量的缺陷倾向性.此外,该类技术还被广泛应用于改进其它软件工程领域中的研究任务.Mou 等人[22]提出了一个基于树的 CNN 模型来提取 AST 的结构信息用于按功能对程序进行分类.Wang 等人[23]采用深度信念网络(DBN, deep brief network)直接从 AST 节点的 token 序列中自动学习语义特征用于缺陷预测.Li 等人[38]提出基于 CNN 的缺陷预测框架 DP-CNN,该方法通过将由 token 序列编码的实值向量与传统代码静态度量相结合,以实现更精确的缺陷预测性能.与上述方法相比,SWIPER 框架利用基于 CFG 路径的细粒度代码表示作为 DNN 的输入.同时,通过使用 CodeBERT 模型在大型语料库上预训练的学习参数,SWIPER 框架能够学习到更加通用的 CFG 路径 token 语义表示,使之可以直接应用于警报确认任务.

### 1.3　基于迁移学习的跨项目警报自动确认技术

现有的静态警报确认模型大多集中于解决来自于同一项目内的警报自动确认问题.然而,在实际的软件开发过程中,需要进行警报确认的项目通常是一个新项目或者缺乏历史数据,往往难以构建有效且实用的警报确认模型.因此,研究人员提出基于迁移学习的方法来改善跨项目场景下警报确认模型性能不佳的问题.Zhang 等人[39]提出一种基于特征排序与匹配的两阶段迁移学习框架 FRM-TL,该方法首先利用特征排序技术从初始特征集合中选择能够满足最优模型性能的特征子空间,然后通过计算源项目和目标项目之间数据分布相似性以构造迁移特征空间用于模型训练.类似地,研究人员也提出许多迁移学习方法来提升跨项目缺陷预测模型的性能[40-42].Zimmermann 等人[43]对跨项目缺陷预测任务的可行性进行了大规模的实证研究.同时,还利用决策树训练基于代码度量和上下文信息的模型用于判断哪种源项目适合于目标项目的缺陷预测.Turhan 等人[44]提出基

于最近邻过滤器的迁移学习方法 NNFilter,该方法首先对代码静态度量进行对数变换,然后基于 k-最近邻算法对可用的训练数据应用相关性过滤器,并选择目标数据中每个实例的 k 个最近实例用于构建预测模型.Nam 等人[45]通过数据归一化来优化迁移成分分析(TCA, transfer component analysis)技术,并提出一种扩展的迁移学习模型 TCA+为源项目和目标项目寻找共享特征空间,使得源项目与目标项目的数据分布在这个空间中是相似的.Qiu 等人[46]提出一种基于 CNN 的可迁移混合特征学习框架 CNN-THFL 来进行跨项目缺陷预测.该方法首先利用 CNN 从 AST 节点的 token 向量中挖掘基于深度学习的语义特征.接着通过应用 TCA 技术从深度学习生成的语义特征和人工设计的静态度量特征中学习可迁移的混合特征.最后,将 CNN-THFL 生成的特征输入分类器以训练缺陷预测模型.本文所提出的方法使用基于频率相似性的检索技术来选择训练实例,并通过预训练技术来学习 CFG 路径 token 的通用表征,从而降低不同软件项目间的数据差异以缓解跨项目场景下警报确认模型精度低的问题.同时,本文还使用微调技术来进一步优化预训练的参数,使其更适配于警报确认任务.

## 2 静态警报自动确认方法 SWIPER

如图 2 所示,本文所提出的静态警报自动确认方法 SWIPER 的整体框架主要包括四个阶段.

① **数据处理:** 解析源代码文件以提取基于 CFG 路径的 token 序列;

② **实例选择:** 基于频率相似性来检索源项目中与目标项目相关的警报实例;

③ **模型训练:** 微调预训练语言模型 CodeBERT 以自动学习基于路径 token 序列的语义表示;

④ **警报确认:** 利用路径语义表示训练分类器以确认静态警报.

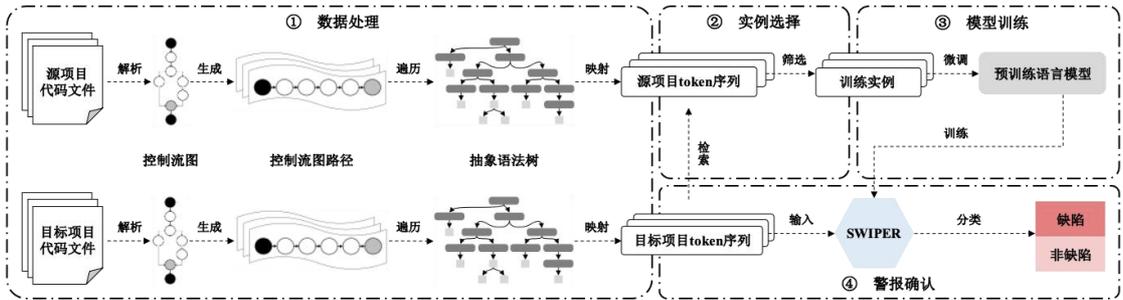

图 2　SWIPER 框架概述

### 2.1　数据处理

在数据处理阶段,SWIPER 框架首先通过静态分析工具 DTS[3](defect testing system)将源代码文件解析为 CFG,解析后的 CFG 中的每个节点可以对应为源代码文件中的一行代码.接着,SWIPER 框架利用面向目标的路径生成算法[39]生成所需的 CFG 路径.该算法的输出是可达过程内执行路径,包括从给定 CFG 的入口节点到目标节点(即包含潜在缺陷信息的警报 IP 处)之间的 CFG 节点序列.针对所生成的每条路径,SWIPER 框架选取两种类型的路径节点进行 token 提取.第一类节点是所有包含警报 IP 处以潜在缺陷的变量(本文中称为 IP 变量)的路径节点.第二类节点是所有包含对 IP 变量存在影响(如定值-引用关系)的其它变量的路径节点.每个路径节点及其包含的原子指令都可以进一步与对应的一个或多个 AST 节点相匹配.表 1 列出了本文所选择的 AST 节点类型及保留的原子指令信息.最后,本方法采用中序遍历策略对生成路径对应的 AST 节点进行遍历.该策略能够使提取的 token 顺序更接近开发人员对代码的阅读习惯.经过上述步骤后,每个静态警报都能被映射为一个包含节点类型和原子指令的 token 序列.

表 1　选择的 AST 节点类别

| 节点类别 | 节点类型 | 原子指令 |
| --- | --- | --- |
| Declaration Node | VariableDeclarator | 变量类型、变量名、常量值及表达式信息 |
| Assignment Node | AssignmentStatement | 变量类型、变量名、常量值及表达式信息 |
| Method Invocation Node | MethodInvocation | 修饰符、函数名、参数类型及参数名信息 |
| Control-Flow Node | IfSelection | 谓词表达式信息 |

| | ForLoop | 循环条件表达式信息 |
| | WhileLoop | 循环条件表达式信息 |
| Jump Node | BreakStatement | - |
| | ContinueStatement | - |
| | ReturnStatement | 变量类型、变量名、常量值及表达式信息 |
| | AssertStatement | 参数类型及参数名信息 |

具体来说,本文选择五种 AST 节点类别作为抽象化的 token,分别是声明语句节点、赋值语句节点、函数调用语句节点、控制流语句节点以及跳转语句节点.而映射就是将原始代码抽象为对应的 token 表示的过程.鉴于函数名和变量名通常是项目特定的,即相同的函数名在不同的项目中实现的功能不一致.因此,本方法没有使用具体的函数名作为 token,而是使用两种函数调用类型(即库函数调用和用户自定义函数调用)作为 token 来表示函数调用语句节点,从而可以在保留部分语义信息的情况下减少数据噪声的出现.而对于代码中出现的变量的抽象(IP 变量除外,IP 变量将被统一抽象为 VariableIP),本方法使用一个整数索引(索引范围从 1 到 N)来区分不同的变量.其中 N 是每段代码中出现的变量的总数,每个变量对应的索引值取决于它在源代码中出现的顺序.例如,第一个出现的变量将被抽象为 Variable1.若该变量在代码中出现多次,那么它们将被赋予相同的索引值.图 3 展示了将图 1 中函数 bad 的代码抽象化为 token 序列的过程.通过将包含潜在缺陷的路径节点 $n_2$ 作为路径生成算法的目标节点,可以生成一条可达过程内路径 $p = n_0 \to n_1 \to n_2$.在本例中,将警报 IP 处导致空指针解引用缺陷的变量 twoInts 被视为 IP 变量.接着按照本节所述方法提取基于 CFG 路径的抽象化 token 序列.与传统的语句级 AST 节点 token 序列[VariableDeclarator, IfStatement, MethodInvocation]相比,本文所提出的基于 CFG 路径的 token 序列包含细粒度的原子指令,能够将导致潜在缺陷的语义信息反馈给后续的 DNN 模型,从而引导模型学习并区分缺陷代码与假阳性警报代码之间细微的差异.

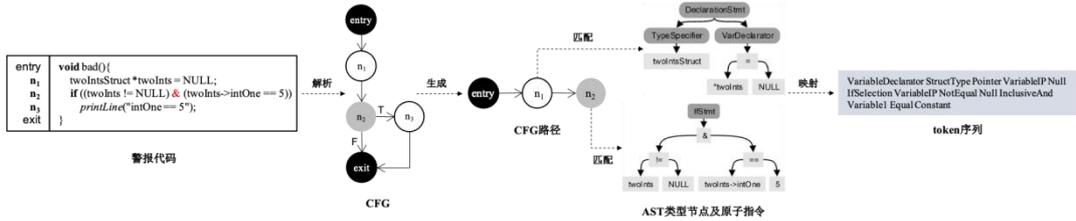

图 3 基于路径的 token 序列提取示例

## 2.2 实例选择

跨项目场景下的警报确认方法通常利用源项目的历史警报数据训练模型,并在缺乏标注数据的目标项目上进行预测.然而,由于不同的软件项目在功能、使用的编程语言以及开发过程中涉及到的开发人员等方面都不尽相同,源项目与目标项目之间通常存在数据分布差异,因此可能出现负迁移问题,导致模型的预测精度较低.为了解决上述问题,本文提出基于频率相似性的实例选择方法来筛选源项目中与目标项目相关的警报实例用于后续模型训练.本文假设不同项目之间的缺陷代码及假阳性警报代码存在相似的代码模式,因此SWIPER 框架利用信息检索领域中基于概率模型的评分算法 BM25[47]来分别计算目标项目中每个警报实例的token 序列与源项目中历史警报实例的 token 序列的相关性分数.如式(1)所示,BM25 算法是一种基于词频和文档长度的统计方法,用于计算查询 q 与文档 D 的匹配度 score(q, D).在本文的任务场景下,BM25 算法首先对每个查询 q(即目标项目中每个警报实例的 token 序列)进行切分解析,生成语素 $q_i$(即 token).然后,对于每个文档 D(即源项目中历史警报实例的 token 序列),计算每个语素$q_i$与 D 的相关性分数.最后,将 $q_i$相对于 D 的相关性分数进行加权求和,从而得到 q 与 D 的相关性分数.式(1)中,t 表示 q 中包含的 token 数量,IDF($q_i$)表示$q_i$的逆文档频率,f($q_i$, D)表示$q_i$在 D 中出现的次数,$L_D$表示 D 的长度(即序列中包含的 token 数量),$L_{avg}$表示源项目中所有历史警报实例的平均长度,$k_1$和 b 是可调参数.通过将目标项目中每个警报实例与源项目中所有历史警报实例所计算得到的相关性分数进行排序后,本文为每个目标项目的警报实例从源项目中筛选前 $n$ 个具有相同类别标签的历史警报实例.为了避免选择与目标项目实例相似但标签矛盾的源项目实例,本文借鉴绝大多数投票法的思想,通过统计源项目中相关性分数排名前 2n+1 个实例中占大多数的类别标签,然后从该类别标签中选择前 n 个警报实例作为训练数据.

$$score(q, D) = \sum_{i=1}^{t} IDF(q_i) \times \frac{f(q_i, D) \times (k_1 + 1)}{f(q_i, D) + k_1 \times (1 - b + b \times \frac{L_D}{L_{avg}})} \tag{1}$$

## 2.3 模型训练

在自然语言处理(NLP, natural language processing)任务中,通常需要对文本进行编码使之能够输入到 DNN 中进行语义表示学习.传统方法大多采用词嵌入技术(如 word2vec[48])将输入序列中的每个词映射到低维的稠密空间中,且语义相近的词向量在空间中离得比较近.但是传统的词嵌入模型在模型在训练完成之后词和向量是一一对应的关系,因此多义词问题无法解决,也无法针对特定任务做动态优化.本文所提出的 SWIPER 框架通过引入预训练语言模型 CodeBERT 来自动学习抽象化后的 CFG 路径的语义表示.CodeBERT 遵循 BERT[49]的模型架构,并使用多层双向的 Transformer[50]作为模型组件.与传统词嵌入方法相比,CodeBERT 利用大规模双模态语料库进行预训练以学习到通用语义表示,有助于完成不同的下游任务.而且经过预训练技术更好地对模型进行初始化,从而在下游任务中获得良好的泛化性能.此外,预训练技术也可以视为一种正则化方法,以避免在规模较小的数据集上出现过拟合的问题.为了保证预训练得到的语义表示能够应用于下游任务(即警报确认),SWIPER 框架利用经过筛选的历史警报数据对模型参数进行微调,从而将 CodeBERT 应用到警报确认任务中.图 4 展示了 SWIPER 框架的整体模型训练过程,在下面的小节中将详细介绍所提出的模型.

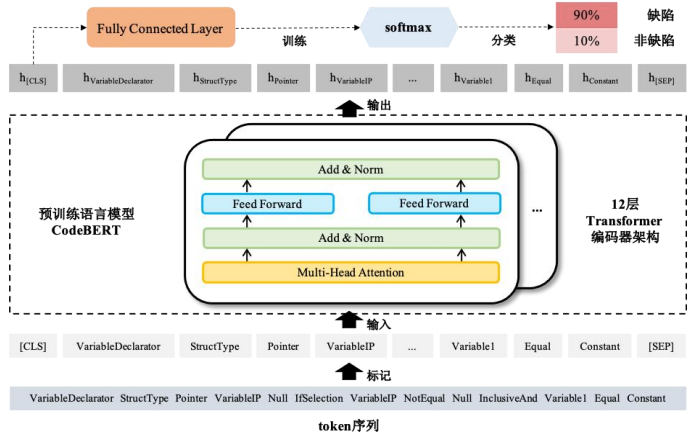

图 4 模型训练过程概述

### 2.3.1 模型架构

如图 4 所示,给定一段抽象警报代码 token 序列,SWIPER 框架采用一个堆叠 12 层 Transformer 编码器架构的预训练语言模型 CodeBERT 来对输入的 token 序列进行编码,使每个 token 获得对应的上下文语义向量表示.每层 Transformer 编码器包括一个多头自注意力层(multi-head self-attention layer)和一个全连接前馈神经网络层(fully-connected feed-forward network layer).这两个子层通过残差连接以防止网络退化,接着对每一层的激活值进行层归一化(layer normalization),最后会在每个位置输出一个对应的隐藏层向量.如图 4 所示,输入的 token 序列中 Constant 对应的隐藏层向量就是 $h_{Constant}$.各层的输出 Output 如式(2)所示,其中 X 表示每一子层的输入向量,LayerNorm 是对隐藏层做层归一化操作,SubLayer 指的是在进行层归一化操作之前的全连接前馈神经网络层或多头自注意力层.多头自注意力层是 Transformer 架构的关键,通过自注意力机制将输入序列中不同位置的 token 信息相关联.多头自注意力层将输入 X 进行线性变换得到矩阵 Q(查询)、K(键值)和 V(值),最终获得自注意力的输出 A 如式(3)所示,其中 $d_k$ 为矩阵的列数,即向量维度.式(3)中计算矩阵 Q 和 K 每一行向量的内积,通过除以 $d_k$ 的平方根进行缩放以防止内积过大.

$$Output = LayerNorm(X + SubLayer(X)) \tag{2}$$

$$A(Q, K, V) = softmax\left(\frac{QK^T}{\sqrt{d_k}}\right)V \tag{3}$$

### 2.3.2 输入表示

对于一段给定的抽象警报代码 token 序列,SWIPER 框架首先将其标记化为预训练语言模型能够接受的输入序列 x = ([CLS], VariableDeclartor, ... , Constant, [SEP]).如图 5 所示,对于一个输入序列中的 token,它的向量表示是由相应的令牌编码(TE, token embedding)、段编码(SE, segment embedding)和位置编码(PE, position embedding)进行相加计算.

• 对于令牌编码,首先随机初始化其向量矩阵,并在训练过程中进行动态调整.特别地,模型在每个输入序列中定义了两个特殊标记[CLS]和[SEP].其中[CLS]出现在输入序列的开头位置,用于引导模型理解当前进行的是分类任务.[CLS]本身并没有语义,经过 12 层自注意力机制计算后得到的是输入序列中其它所有 token 语义的加权平均.相比于其它本身带有语义的 token,[CLS]可以更好地表征整个输入序列的语义.因此与其相对应的最终隐藏层输出向量 h[CLS]可以作为分类任务的聚合序列表示.而[SEP]则是用于区分属于不同序列的 token.本文所使用的警报确认数据集中只有单模态的输入序列(即警报代码的 token 序列),因此[SEP]被添加到序列的末尾.

• 对于段编码,其向量矩阵也是被随机初始化的.由于所有 token 都是属于同一个序列,因此所有 token 的输入表示都需要加上一个向量 $E_A$.

• 对于位置编码,通过使用学习到的向量矩阵来表示序列的顺序.需要注意的是,CodeBERT 模型限制输入序列的最大长度为 512.因此,若输入序列中的 token 长度小于 512,则需要输入序列末尾使用标记[PAD]来填充.若输入序列中的 token 长度大于 512,则需要进行截断.

图 5　输入表示

### 2.3.3 预训练及微调

为了学习输入序列中 token 的语义表示,本文利用预训练语言模型 CodeBERT 的学习参数对 SWIPER 框架进行初始化.SWIPER 框架的目标是对给定的静态警报相关代码进行自动分类,以判断代码是否包含潜在缺陷.通过使用 CodeBERT 在大型语料库上预训练的学习参数,SWIPER 框架能够学习到更加通用的 token 语义表示,使之可以直接应用于静态警报自动确认任务.同时,学习到的通用表征可以降低不同软件项目间的数据差异,从而缓解跨项目场景下警报确认模型精度低的问题.在深度学习领域,微调技术可以优化预训练的参数,使其更适配于下游任务.针对本文中的警报确认任务,SWIPER 框架将其与文本分类任务进行类比,其中模型的输入是一段警报相关代码的 token 序列,输出是预期的标签.微调过程使用经过筛选的历史警报数据集作为训练语料对 CodeBERT 模型进行有监督的训练,从而改变原始 CodeBERT 模型中的权重矩阵.具体来说,语料库 C 中的每个实例可以表示为一个二元组 $C_i$ = {text, label},其中 text 表示代码 token 序列,label 表示人工标ésの警报类别标签.微调的目的是通过学习映射 text → label 的条件概率分布 p(label | text),从而最小化交叉熵损失函数.

### 2.4 警报确认

如图 4 所示,为了将训练好的模型应用于静态警报的分类任务中,SWIPER 框架在[CLS]标记的最终隐藏层输出 h[CLS]后增加一个全连接层(fully connected layer)作为分类层,并利用 softmax 函数来分别计算两个分类标签(即缺陷或非缺陷)的概率,最后概率较大值所对应的标签即为模型分类的结果.

## 3　实验设置

### 3.1　实验数据集

如表 2 所示,本文在 8 个 C 语言开源项目的公开警报数据集上进行实验.在这些数据集中,有缺陷成分的百分比为 44.5%~87.3%.其中,前 3 个项目(C Test Suite、ITC 和 Juliet)来自于美国国家标准与技术研究院发布的美国国家公开漏洞数据库 SARD(software assurance reference dataset).项目中包含大量常见缺陷的测试用例,已被

广泛应用于评估 ASAT 的性能[51-53].此外,本文还从已有研究[39,54,55]中选择了 5 个来自开源软件项目的公开数据集.每个数据集包含在真实软件开发实践中通过 ASAT 检测出的警报实例,且由专业开发人员进行人工审查得到标签结果.这些项目的代码行数(LoC, lines of code)从 2K 到 79K 不等,且功能多样.第 3 列则展示了项目中由 ASAT 所报告的警报数量,范围从 57 到 1278.本文使用上述研究论文或开源数据库中报告的人工审查结果作为数据集的标签以保证可靠性.

表 2　实验数据集

| 项目 | LoC | 警报数量 | 缺陷率(%) |
|---|---|---|---|
| C Test Suite | 36134 | 146 | 46.8 |
| ITC | 42110 | 290 | 50.0 |
| Juliet | 79324 | 1278 | 44.5 |
| antiword-0.37 | 20213 | 59 | 59.3 |
| barcode-0.98 | 3409 | 63 | 87.3 |
| spell-1.0 | 1991 | 57 | 54.4 |
| sphinxbase-0.2 | 22517 | 220 | 50.0 |
| uucp-1.07 | 52595 | 759 | 71.0 |

**3.2　实验评估指标**

为了合理评估警报确认模型的分类结果,本文采用两个常用的度量指标精确率和召回率来衡量模型的预测性能.对于一个二分类问题,通过将模型预测的结果和人工标注的标签相匹配,可以构建相应的二维混淆矩阵,混淆矩阵中的四个值分别表示:1)将真实缺陷预测为真实缺陷的实例数量($N_{buggy \rightarrow buggy}$);2)将真实缺陷预测为假阳性警报的实例数量($N_{buggy \rightarrow clean}$);3)将假阳性警报预测为真实缺陷的实例数量($N_{clean \rightarrow buggy}$);4)将假阳性警报预测为假阳性警报的实例数量($N_{clean \rightarrow clean}$).如式(4)和(5)所示,本文计算假阳性警报类别的精确率(Precision)和召回率(Recall).两个度量指标的取值范围在 0 到 1 之间,值越接近 1 表示模型性能越好.具体地说,精确率越高表示模型所过滤的警报中真实的假阳性警报占比越高(即漏报率越低),召回率越高表示模型所过滤的假阳性警报占所有假阳性警报的比例越高(即误报率越低).

$$Precision = \frac{N_{clean \rightarrow clean}}{N_{clean \rightarrow clean} + N_{buggy \rightarrow clean}} \tag{4}$$

$$Recall = \frac{N_{clean \rightarrow clean}}{N_{clean \rightarrow clean} + N_{clean \rightarrow buggy}} \tag{5}$$

**3.3　实验设计**

针对跨项目场景下的警报确认(CPWI, cross-project warning identification)任务,本文从 8 个项目中选择一个项目作为源项目,接着依次使用剩余的项目作为目标项目进行 CPWI 任务.例如,当表 2 中的 spell 作为源项目时,存在 7 个 CPWI 项目组合,因此总共有 56 种可能的项目组合.对于每个项目组合,利用 2.2 节中所提出的方法从源项目中筛选出与目标项目相关的历史警报实例作为训练集用于 SWIPER 框架中模型的训练及微调.然后利用基于源项目的历史警报数据训练的 softmax 分类器对目标项目中的警报进行确认.为了评估 SWIPER 框架在 CPWI 任务中的性能,本文将所提出的方法与下面 6 个与本文相关的基线方法进行比较.

● FRM-TL[39]:一种用于 CPWI 任务的两阶段迁移学习框架,通过利用 PVC(path-variable characteristic)特征进行排序与匹配来减小项目间的数据分布差异.

● TCA[45]:一种使用基于传统代码静态度量的 CM(common metric)特征的经典迁移学习方法,通过最小化源项目与目标项目的特征空间距离来达到两个项目之间数据分布相似的目的.表 3 列出了本文用于构建分类器的 28 个 CM 特征,所列出的 CM 特征可在 Menzies 等人[17]的论文中找到完整定义.

● NNFilter[44]:一种基于欧氏距离的模型,使用最近邻分类器选择源项目中与目标项目有类似分布的数据.

● DBN[23]:一种利用深度学习模型 DBN 来自动学习 AST 节点序列的语义特征用于缺陷预测的方法.

● DP-CNN[38]:一种将 CNN 生成的基于 AST 的语义特征向量与传统的 CM 特征向量相结合构建缺陷预测模型的方法.

● CNN-THFL[46]:一种基于 CNN 的可迁移混合特征学习框架,通过在 DP-CNN 方法的基础上引入 TCA 技

术学习迁移后的特征来构建跨项目场景下的缺陷预测模型.

表 3   传统代码静态度量特征

| 类别 | 特征名 |
|---|---|
| LOC | total_loc, blank_loc, comment_loc, executable_loc |
| McCabe | cyclomatic_complexity, design_complexity |
| Halstead | unique_operands, unique_operators, total_operands, total_operators, halstead_vocabulary, halstead_length, halstead_volume, halstead_level, halstead_difficulty, halstead_effort, halstead_error, halstead_time |
| Miscellaneous | branch_count, decision_count, call_pairs, condition_count, multiple_condition_count, cyclomatic_density, decision_density, design_density, norm_cyclomatic_complexity, formal_parameters |

对于项目内场景下的警报确认(WPWI, within-project warning identification)任务,本文将每个项目的数据集分为训练集和测试集,并在警报确认模型中使用两折交叉验证来评估分类的效果.具体来说,针对来自同一个项目的历史警报,将数据集随机划分为两个大小相等的子集,利用其中一个子集作为训练集用于 SWIPER 框架中模型的训练及微调,而另一个子集的数据则用于测试,然后再交换两个子集进行一次重复实验.为了评估 SWIPER 框架对 WPWI 任务中的性能,本文将所提出的方法与 3 个常用的机器学习分类器进行比较.考虑到机器学习算法的流行性和多样性,本文分别选择了逻辑回归分类器 LR(logistic regression)、基于决策树算法的分类器 J48 以及支持向量机分类器 SVM(support vector machine)来作为基线方法.上述机器学习分类器均使用基于 CM 特征的数据集来训练 WPWI 模型.

本文利用开源工具 Prest[56]来收集 CM 特征用于实验分析.针对 FRM-TL、TCA、NNFilter 以及 CNN-THFL 这四个基线方法的模型实现,本文利用作者提供开源代码来复现相关模型.针对 DBN 以及 DP-CNN 这两个基线方法,本文通过已有论文中描述的相同的网络结构以及参数来复现相关模型.针对 LR、J48 以及 SVM 这三个基线方法,本文利用开源平台 Weka[57]中实现的模型进行实验.本文所提出的 SWIPER 框架利用 Huggingface 平台的"codebert-base"版本作为基础模型架构,由 12 层的 Transformer 编码器组成.SWIPER 框架的具体参数设置如下: 源项目与目标项目的最大序列长度为 512、批处理大小(batch size)为 16、学习率(learning rate)为 1e-6 以及训练的 epoch 设置为 15.在每个 epoch 中,模型从训练集中随机选择 20%的数据进行验证.为了防止过拟合问题,本文采用早停(early stopping)机制.若模型在验证集上超过 5 个 epoch 没有性能改善,则训练过程停止.在测试阶段,将训练好的模型在测试集上进行模型性能验证,并报告 3.2 节中所提出的评估指标以进行比较实验.为了避免数据选择的随机性带来的影响[58],本文对每组实验都独立重复 10 次取平均的评估指标结果.本文中的所有实验均在装有 4 块 Nvidia GTX 1080Ti GPU 的服务器上进行.

本文采用 Wilcoxon 符号秩检验方法来验证所提出的 SWIPER 框架与基线方法的在警报确认性能上是否有显著差异.Wilcoxon 符号秩检验是一种非参数检验,意味着不需要假设总体数据满足正态分布.当方法的 $p$ 值小于 0.05 时,说明两个模型的性能差异在统计意义上具有显著性.此外,本文使用 Scott-Knott Effect Size Difference(ESD)检验方法从 3.2 节中所提出的两个评估指标的角度对模型进行排名.Scott-Knott ESD 是一种均值比较方法,它利用层次聚类将数据集划分为统计学上差异不可忽略的不同的组[59].在进行 Scott-Knott ESD 检验的同时,本文还计算 Cohen 效应量 $d$ 来衡量 SWIPER 框架与基线方法之间性能的差异在实际应用中是否重要.当 $|d| \leq 0.2$ 时,认为是可以忽略的(N, negligible);当 $0.2 < |d| \leq 0.5$ 时,认为是小的(S, small);当 $0.5 < |d| \leq 0.8$ 时,认为是中等的(M, medium);当 $|d| \geq 0.8$ 时,认为是大的(L, large).如果一个模型对于另一个模型的效应量 $d$ 为正且大于 0.2,则表示该模型更具有使用价值.

## 4   实验结果与分析

为了评估基于路径语义表示的静态警报自动确认方法 SWIPER 的有效性,本文研究了以下三个研究问题(RQ, research question).

- **RQ1**: 与其它基线方法相比,SWIPER 能否获得更好的 CPWI 结果?
- **RQ2**: 与其它基线方法相比,SWIPER 能否获得更好的 WPWI 结果?
- **RQ3**: 基于路径的语义表示以及预训练语言模型对警报确认任务是否有效?

**4.1 RQ1实验结果与分析**

表 4 和表 5 分别列出了在每个目标项目上使用本文所提出的方法 SWIPER 以及 6 个基线方法时的平均 Precision 和 Recall 结果(按行显示).每个目标项目行中的值都是 70 次实验结果(7个项目组合×10 次独立重复实验)的平均值.如表 4 和表 5 所示,在每一个目标项目行的对比实验中,SWIPER 与所有基线方法之中最好的结果会被加粗表示.表 4 和表 5 的最后三行分别表示每个方法在所有目标项目上性能结果的平均值(见 **Avg.**行)、每个基线方法与 SWIPER 进行 Wilcoxon 符号秩检验的 $p$ 值(见 **$p$-value** 行)以及每个基线方法与 SWIPER 之间的 Cohen 效应量大小(见 **Cohen's $d$** 行).图 6 则展示了对所有方法进行 Scott-Knott ESD 检验的排名结果,同种颜色的方法属于同一个排名类别.基于对表 4、表 5 以及图 6 的观察,可以得出以下分析结果.

- 在 Scott-Knott ESD 检验排名中,SWIPER 在两个评估指标中都排名第一,说明利用 SWIPER 方法所构建的 CPWI 模型对比其它基线方法具有显著的性能提升.

- 针对精确率来说,SWIPER 在所有目标项目上取得的平均 Precision 值为59.1%.与其它 6 个基线方法相比较,平均 Precision 值的提升幅度在 8.3%~16.1%之间.而从召回率的结果来看,SWIPER 在所有目标项目上取得的平均 Recall 值为 75.5%.与其它 6 个基线方法相比较,平均 Recall 值的提升幅度在 19.0%~46.2%之间.

- 根据 Wilcoxon 符号秩检验的结果,SWIPER 在两个评估指标中对于所有基线方法的 p 均小于 0.05,说明 SWIPER 和基线方法的模型性能差异在统计意义上具有显著性.根据 Cohen 效应量大小的结果,SWIPER 在两个评估指标中相对于其它基线方法的效应量 $d$ 均为正且大于 0.5(即性能差异在实际应用中中等偏大),这也充分表明跨项目场景下 SWIPER 框架的实用性.

表 4  SWIPER 与其它基线方法在 CPWI 任务中的 Precision 指标比较结果

| 目标项目 | FRM-TL | TCA | NNFilter | DBN | DP-CNN | CNN-THFL | SWIPER |
|---|---|---|---|---|---|---|---|
| C Test Suite | 52.5% | 36.9% | 51.8% | 52.5% | 52.2% | 53.5% | **57.6%** |
| ITC | 49.7% | 50.5% | 52.7% | 50.6% | 51.0% | 52.3% | **53.5%** |
| Juliet | 49.8% | **79.3%** | 56.8% | 58.3% | 59.9% | 68.2% | 58.1% |
| antiword | 36.4% | 29.5% | 44.3% | 47.4% | 46.1% | 37.7% | **59.0%** |
| barcode | 20.5% | 20.4% | 39.4% | 16.0% | 14.9% | 18.1% | **63.2%** |
| spell | **62.0%** | 46.5% | 60.0% | 50.6% | 56.8% | 49.9% | 51.9% |
| sphinxbase | 43.2% | 60.4% | 55.1% | 54.1% | 55.1% | 61.1% | **68.5%** |
| uucp | 29.8% | 22.0% | 46.4% | 32.4% | 34.6% | 37.7% | **61.0%** |
| **Avg.** | 43.0% | 43.2% | 50.8% | 45.2% | 46.3% | 47.3% | **59.1%** |
| *p*-value | $2.10\times10^{-14}$ | $1.22\times10^{-10}$ | $3.73\times10^{-6}$ | $1.21\times10^{-15}$ | $1.09\times10^{-12}$ | $1.97\times10^{-8}$ | - |
| **Cohen's $d$** | 1.072 (L) | 0.884 (L) | 0.555 (M) | 1.156 (L) | 0.982 (L) | 0.793 (M) | - |

表 5  SWIPER 与其它基线方法在 CPWI 任务中的 Recall 指标比较结果

| 目标项目 | FRM-TL | TCA | NNFilter | DBN | DP-CNN | CNN-THFL | SWIPER |
|---|---|---|---|---|---|---|---|
| C Test Suite | 29.2% | 36.6% | 44.6% | 49.7% | 43.9% | 40.0% | **74.3%** |
| ITC | 31.9% | 60.5% | 64.6% | 52.7% | 47.6% | 43.8% | **72.7%** |
| Juliet | 19.3% | 47.6% | 51.8% | 57.3% | 46.6% | 43.9% | **93.7%** |
| antiword | 31.0% | 38.2% | 51.8% | 61.8% | 51.9% | 33.8% | **96.7%** |
| barcode | 44.6% | 43.8% | 39.3% | **62.9%** | 51.4% | 58.2% | 40.0% |
| spell | 29.1% | 52.3% | 45.1% | 56.2% | 50.8% | 47.1% | **91.9%** |
| sphinxbase | 20.5% | 51.1% | 31.8% | 56.3% | 49.1% | 42.0% | **86.0%** |
| uucp | 28.6% | 47.2% | 26.2% | 55.0% | **57.4%** | 47.8% | 48.9% |
| **Avg.** | 29.3% | 47.2% | 44.4% | 56.5% | 49.8% | 44.6% | **75.5%** |
| *p*-value | $2.98\times10^{-21}$ | $1.12\times10^{-15}$ | $3.76\times10^{-14}$ | $9.26\times10^{-14}$ | $2.86\times10^{-13}$ | $2.58\times10^{-15}$ | - |
| **Cohen's $d$** | 2.223 (L) | 1.299 (L) | 1.207 (L) | 1.156 (L) | 1.258 (L) | 1.448 (L) | - |

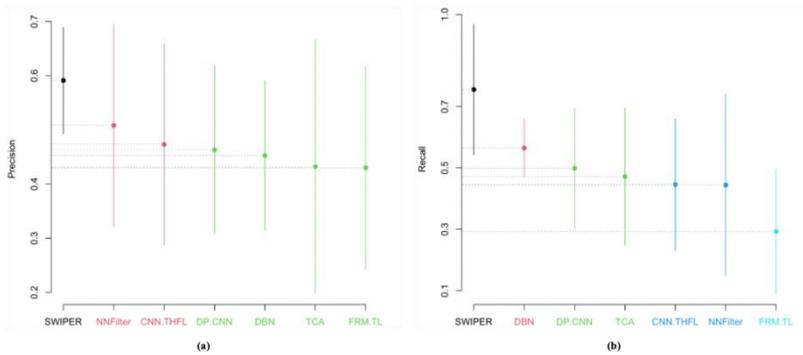

图 6　RQ1 的 Scott-Knott ESD 排名结果. (a) Precision 评估指标; (b) Recall 评估指标.

**4.2　RQ2实验结果与分析**

表 6 中列出了每个目标项目在 SWIPER 和 3 个基线方法上训练 WPWI 模型时的平均 Precision 和 Recall 结果.每个目标项目行中的值是重复 10 次二折交叉验证(即每个模型都有 20 次实验结果)的平均值.如表6所示,在每个目标项目行的对比实验中,SWIPER 与其它基线方法之中更好的结果会被加粗表示.图 7 则展示了对所有方法进行 Scott-Knott ESD 检验的排名结果.基于对表 6 以及图 7 的观察,可以得出以下分析结果.

• SWIPER 在两个评估指标下的 Scott-Knott ESD 检验排名中均优于其它基线方法,因此在 WPWI 任务中使用基于路径语义表示的 SWIPER 方法也能得到显著的性能提升.

• SWIPER 在所有被测项目上取得的平均 Precision 和 Recall 值分别为 74.1%和 86.9%.与其它 3 个基线方法相比较,平均 Precision 值的提升幅度在 11.7%~14.1%之间,平均 Recall 值的提升幅度在 29.0%~35.5%之间.

• 与 3 个基线方法相比,本文所提出的方法 SWIPER 有着明显的优势,在绝大多数目标项目数据集上都获得了更好的模型效果.根据 Wilcoxon 符号秩检和 Cohen 效应量大小的结果,SWIPER 相对于所有基线方法在模型性能上有大规模的优势,且性能差异在统计意义上具有显著性.以上结果也表明 SWIPER 框架在项目内场景下的依旧具有实用价值.

表 6　SWIPER 与 3 个基线方法在 WPWI 任务中的评估指标比较结果

| 目标项目 | Precision | | | | Recall | | | |
|---|---|---|---|---|---|---|---|---|
| | LR | J48 | SVM | SWIPER | LR | J48 | SVM | SWIPER |
| C Test Suite | 50.5% | 46.2% | 49.4% | **69.7%** | 43.2% | 48.4% | 45.4% | **100.0%** |
| ITC | 46.8% | 48.6% | 47.7% | **70.3%** | 24.3% | 64.6% | 49.7% | **98.9%** |
| Juliet | 76.4% | 77.1% | 78.2% | **80.9%** | 88.0% | 81.6% | 82.3% | **93.4%** |
| antiword | 66.4% | 76.1% | 71.2% | **77.4%** | 78.0% | 80.6% | 75.2% | **85.7%** |
| barcode | 46.6% | 17.5% | 34.0% | **63.3%** | 8.0% | 5.7% | 11.1% | **66.7%** |
| spell | 69.6% | 61.0% | 67.2% | **71.0%** | 68.2% | 49.5% | 58.7% | **92.7%** |
| sphinxbase | 70.4% | 80.2% | 70.8% | **81.0%** | 68.5% | 75.6% | 64.6% | **91.9%** |
| uucp | 69.5% | 73.2% | **80.9%** | 79.2% | 44.4% | 57.0% | 24.3% | **65.7%** |
| **Avg.** | 62.0% | 60.0% | 62.4% | **74.1%** | 52.8% | 57.9% | 51.4% | **86.9%** |
| *p*-value | $9.95\times10^{-9}$ | $6.10\times10^{-7}$ | $3.42\times10^{-7}$ | - | $4.64\times10^{-12}$ | $1.15\times10^{-10}$ | $7.74\times10^{-13}$ | - |
| **Cohen's *d*** | 0.971 (L) | 0.889 (L) | 0.899 (L) | - | 1.643 (L) | 1.513 (L) | 1.894 (L) | - |

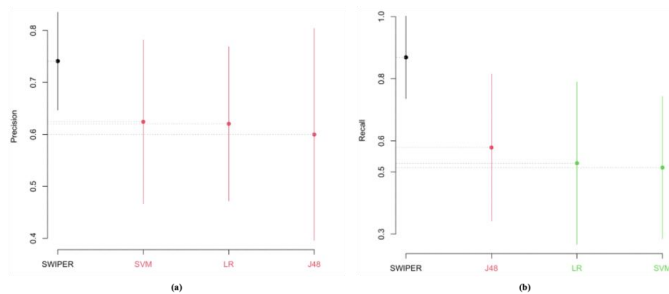

图 7　RQ2 的 Scott-Knott ESD 排名结果. (a) Precision 评估指标; (b) Recall 评估指标.

### 4.3　RQ3实验结果与分析

为了回答 RQ3,本文通过以下两小节的消融实验来验证 SWIPER 方法中不同组件的有效性.

#### 4.3.1　代码表示影响分析

本节的消融实验选择两种传统的代码表示方法来替换 SWIPER 方法中的路径语义表示,并利用预训练语言模型 CodeBERT 来训练基线方法模型在跨项目场景下进行性能比较.具体来说,两个基线方法分别是基于代码静态度量的模型 CodeBERT-CM,以及基于 AST 语义表示的模型 CodeBERT-AST.为了保证实验的公平性,CodeBERT-CM 与 CodeBERT-AST 的训练过程与 3.3 节中所述一致.

表 8 列出了在每个目标项目上使用 SWIPER 和两个基线方法(CodeBERT-CM 与 CodeBERT-AST)时的平均 Precision 和 Recall 结果.在每个目标项目行的对比实验中,SWIPER 与 CodeBERT-CM、CodeBERT-AST 之中最好的评估指标结果会被加粗表示.与传统的代码表示方法相比,基于路径语义的代码表示有着巨大的优势,在绝大多数的目标项目数据集上都获得了最佳的模型性能效果.具体来说,在 8 个目标项目中,本文所提出的方法 SWIPER 在 Precision 和 Recall 指标下都取得了 6 个最佳的结果.针对精确率来说,SWIPER 比两个基线方法分别提升了 7.6% 和 6.1%.而从召回率的结果来看,SWIPER 的提升幅度分别是 17.8% 和 15.9%.根据 Wilcoxon 符号秩检和与 Cohen 效应量大小检验的结果,SWIPER 相对于 CodeBERT-CM 与 CodeBERT-AST 在 Precision 指标下有小幅度到中等的优势,在 Recall 指标下有中等的优势,且模型的性能差异也具有统计显著性.总的来说,基于路径的细粒度语义表示通过有效区分程序代码中细微的差异,可以引导模型更精确地学习与潜在缺陷相关的语义表征,从而提高模型在警报确认任务中的性能.

表 7　SWIPER 与 CodeBERT-CM、CodeBERT-AST 在 CPWI 任务中的评估指标比较结果

| 目标项目 | Precision | | | Recall | | |
|---|---|---|---|---|---|---|
| | CodeBERT-CM | CodeBERT-AST | SWIPER | CodeBERT-CM | CodeBERT-AST | SWIPER |
| C Test Suite | 56.7% | 56.9% | **57.6%** | 52.4% | 60.5% | **74.3%** |
| ITC | 52.8% | 53.2% | **53.5%** | 64.3% | 61.9% | **72.7%** |
| Juliet | **67.5%** | 65.3% | 58.1% | 66.7% | 60.5% | **93.7%** |
| antiword | 47.6% | 48.1% | **59.0%** | 51.1% | 67.0% | **96.7%** |
| barcode | 27.6% | 38.1% | **63.2%** | 55.8% | **58.9%** | 40.0% |
| spell | **53.6%** | **53.6%** | 51.9% | 67.1% | 60.6% | **91.9%** |
| sphinxbase | 64.2% | 65.1% | **68.5%** | 50.5% | 49.3% | **86.0%** |
| uucp | 42.3% | 43.7% | **61.0%** | 53.6% | **58.1%** | 48.9% |
| Avg. | 51.5% | 53.0% | **59.1%** | 57.7% | 59.6% | **75.5%** |
| *p*-value | $4.94 \times 10^{-10}$ | $9.63 \times 10^{-6}$ | - | $2.20 \times 10^{-10}$ | $3.99 \times 10^{-10}$ | - |
| *Cohen's d* | 0.555 (M) | 0.467 (S) | - | 0.615 (M) | 0.577 (M) | - |

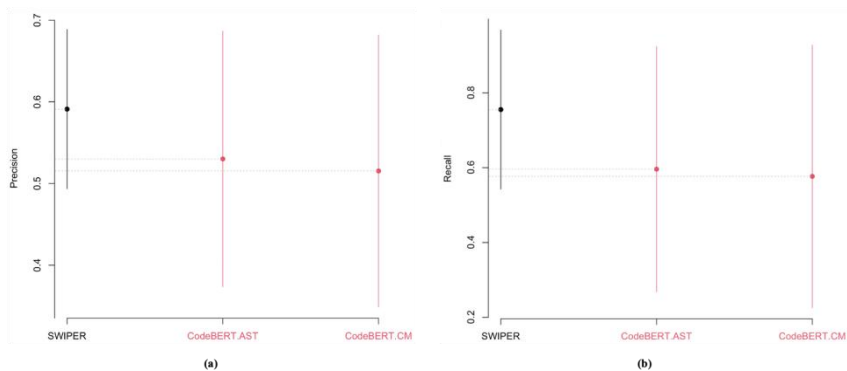

图 8　RQ3 中第一个消融实验的 Scott-Knott ESD 排名结果. (a) Precision 评估指标; (b) Recall 评估指标.

#### 4.3.2　DNN 架构影响分析

本节的消融实验选择已有论文中常用的四种 DNN 架构(CNN、DBN、词嵌入模型 Word2Vec 以及预训练语言模型 BERT)来替换 SWIPER 方法中的预训练语言模型 CodeBERT,并通过训练 CPWI 模型来进行性能对比.也就是说,本节进行对比的四个基线方法分别是利用路径语义表示训练的 CNN 模型(CNN-Path)、DBN 模型(DBN-Path)、Word2Vec 模型(Word2Vec-Path)以及 BERT 模型(BERT-Path).关于 DNN 架构的实现,本文采用已

有论文[16,38,48]中开源的模型代码来进行复现 CNN、DBN 以及 Word2Vec,并利用 Huggingface 平台开源的模型接口"bert-base-cased"来复现 BERT.为了保证实验的公平性,四个基线方法的训练过程与 3.3 节中所述一致.

表 8 和表 9 列出了在每个目标项目上使用 SWIPER 和四个基线方法时的平均 Precision 和 Recall 结果.在每一个目标项目行的对比实验中,SWIPER 与四个基线方法之中最好的评估指标结果会被加粗表示.如图 9 所示,在比较不同的 DNN 架构时,本文所提出的使用预训练语言模型 CodeBERT 的方法 SWIPER 在两个评估指标中的 Scott-Knott ESD 统计检验排名都比四个基线方法要高,这说明通过使用基于自然语言和编程语言的双模态预训练模型 CodeBERT 可以学习到更加通用的代码语义表示以减小不同项目间的数据差异,从而在跨项目场景下能够比传统的深度神经网络架构获得更好的模型性能.与四个基线方法相比较,本为所提出的方法 SWIPER 在平均 Precision 值的提升幅度在 8.9%~13.6% 之间,而在平均 Recall 值的提升幅度在 8.2%~28.6% 之间.同时,根据 $p$-value 和 Cohen's $d$ 的检验结果,SWITER 相对于四个基线方法在绝大多数的目标项目数据集中具有大规模尺度的优势,且性能的差异在统计学上具有显著性.综上,使用预训练语言模型在警报确认任务上可以获得优于传统 DNN 架构的模型性能,并在跨项目场景下相对于现有方法更具有实用性.

**表 8** SWIPER 与四个基线方法在 CPWI 任务中的 Precision 指标比较结果

| 目标项目 | CNN-Path | DBN-Path | Word2Vec-Path | BERT-Path | SWIPER |
|---|---|---|---|---|---|
| C Test Suite | 54.3% | 53.9% | 53.4% | 55.8% | **57.6%** |
| ITC | 52.6% | 51.5% | 52.3% | **54.2%** | 53.5% |
| Juliet | 57.6% | **58.4%** | 58.3% | 57.8% | 58.1% |
| antiword | 45.0% | 48.5% | 55.1% | 55.6% | **59.0%** |
| barcode | 15.5% | 16.0% | 32.5% | 32.4% | **63.2%** |
| spell | **53.4%** | 51.5% | 53.1% | 49.1% | 51.9% |
| sphinxbase | 53.6% | 55.7% | 58.4% | 59.9% | **68.5%** |
| uucp | 32.3% | 32.1% | 35.4% | 36.9% | **61.0%** |
| Avg. | 45.5% | 46.0% | 49.8% | 50.2% | **59.1%** |
| $p$-value | $1.24 \times 10^{-14}$ | $4.47 \times 10^{-14}$ | $1.08 \times 10^{-8}$ | $3.34 \times 10^{-6}$ | - |
| Cohen's $d$ | 1.086 (L) | 1.082 (L) | 0.807 (L) | 0.704 (M) | - |

**表 9** SWIPER 与四个基线方法在 CPWI 任务中的 Recall 指标比较结果

| 目标项目 | CNN-Path | DBN-Path | Word2Vec-Path | BERT-Path | SWIPER |
|---|---|---|---|---|---|
| C Test Suite | 42.2% | 51.6% | 55.7% | 70.6% | **74.3%** |
| ITC | 41.9% | 49.8% | 53.5% | 59.7% | **72.7%** |
| Juliet | 39.7% | 52.3% | 64.1% | 75.6% | **93.7%** |
| antiword | 51.7% | 56.9% | 69.4% | 64.4% | **96.7%** |
| barcode | 54.6% | 61.1% | 48.9% | **72.5%** | 40.0% |
| spell | 45.7% | 54.6% | 61.0% | 77.8% | **91.9%** |
| sphinxbase | 45.6% | 54.8% | 52.7% | 53.0% | **86.0%** |
| uucp | 53.6% | 54.2% | 52.4% | **64.3%** | 48.9% |
| Avg. | 46.9% | 54.4% | 57.2% | 67.3% | **75.5%** |
| $p$-value | $1.47 \times 10^{-16}$ | $6.08 \times 10^{-15}$ | $2.94 \times 10^{-9}$ | $4.48 \times 10^{-4}$ | - |
| Cohen's $d$ | 1.455 (L) | 1.302 (L) | 0.791 (M) | 0.340 (S) | - |

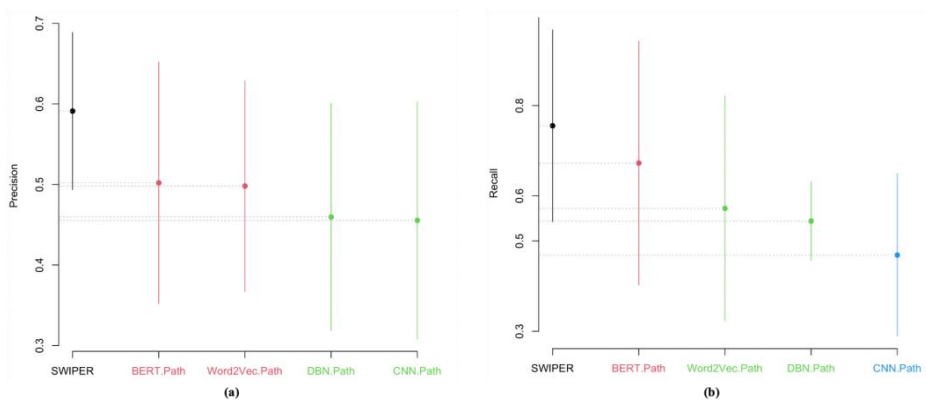

图 9 RQ3 中第二个消融实验的 Scott-Knott ESD 排名结果. (a) Precision 评估指标; (b) Recall 评估指标.

#### 4.4　有效性威胁分析

本文从以下三方面对所提出方法 SWIPER 的有效性威胁进行分析.

* 外部有效性：数据集的质量是本文中对外部有效性的主要威胁.由于人工审查成本较大,目前还没有公开可用的基准数据集中包含由 ASAT 报告的带有标签的实际软件项目的历史警报数据,因此训练高精度的警报确认模型是一大挑战.本文选择了来自 SARD 漏洞库中人工合成的相关基准测试套件以及来自开源软件项目的真实警报数据集作为本文的基准数据集.在这些项目之外训练本文所提出的 SWIPER 模型可能会得到更好或更差的结果,因此构建具有普遍性的警报数据语料库将是未来考虑的工作之一.

* 内部有效性：由于计算资源的限制,本文并未对模型进行全面的超参数优化.本文采用已有研究建议的最佳架构进行模型构建,因此,通过对参数进行额外的调优可能会对模型的性能进行改善.其次,本文在数据处理阶段的路径选择上具有随机性.因此,下一步研究工作将考虑根据覆盖准则(如路径覆盖)来生成给定函数的路径集合,通过融合集合中不同路径上与潜在缺陷相关的语义特征来进一步提升本文所提出的方法的普适性.

* 结构有效性：首先,本文选择两个常用的度量指标精确率和召回率来衡量模型的性能.而实际上还有其它评估指标(如 G-measure)也可以在未来的工作中使用以验证模型的性能.其次,本文选择以 C 语言程序编写的软件项目作为基准数据集来证明所提方法的有效性.由于大多数编程语言中都存在 CFG 和 AST 等抽象表示,因此所提出的方法可以比较容易地通过进行少量修改(如根据不同的编程语言选择不同的 AST 节点类型)在新的编程语言(如 Java)中进行扩展.例如,首先基于已有的面向 Java 编程语言的缺陷数据集(如 PROMISE[60]),构建缺陷数据所对应的开源项目代码语料库.然后利用开源工具(如 PROGEX[61])抽取缺陷所对应的 Java 函数的抽象化代码表示.最后,本文所提出的基于预训练语言模型的训练框架并不依赖于特定的编程语言,因此可以通过类似的序列化输入数据格式构建面向 Java 编程语言的警报自动确认模型.

## 5　总结与展望

静态警报自动确认技术对提高软件测试自动化是非常重要的.本文提出的 SWIPER 方法通过利用预训练语言模型从基于 CFG 路径的代码表示中自动学习语义特征用于警报确认.针对传统特征表示无法区分缺陷导致模型假阳率高的问题,本文利用路径敏感性分析从控制流图中提取与静态警报相关的细粒度路径节点序列,从而引导模型更准确地学习与潜在缺陷相关的语义信息.为了缓解 CPWI 模型精度低的问题,本文通过在警报确认任务上微调预训练语言模型以达到提高模型泛化能力的目的,并利用信息检索领域中基于概率模型的 BM25 算法来筛选源项目中与目标项目相关性分数较高的警报实例用于构建模型训练数据集,从而减小不同项目间的数据分布差异.本文在 8 个开源警报数据集上将 SWIPER 与多种基线方法进行性能对比分析.实验结果验证了所提方法不仅可以有效地学习警报相关代码的语义信息,还可以进一步提升 CPWI 任务的性能.本文采用 CFG 作为程序代码的抽象表示模型,然而 ASAT 检测缺陷的过程与代码结构、数据流和依赖性均有关系.因此,下一步研究工作考虑将不同抽象表示的属性组合在一个联合数据结构中用于提取与缺陷相关的语义特征,从而使警报确认模型更加准确.此外,对 DNN 模型的输出结果进行更多的可解释性分析也具有重要意义.

**References**:

[1]　Ayewah N, Pugh W. The Google FindBugs fixit. In: Proc. of the 19th Int'l Symp. on Software Testing and Analysis. New York: Association for Computing Machinery, 2010. 241–252.

[2]　Bessey A, Block K, Chelf B, Chou A, Fulton B, Hallem S, Henri-Gros C, Kamsky A, McPeak S, Engler DR. A few billion lines of code later: using static analysis to find bugs in the real world. Commun. ACM, 2010,53(2):66–75.

[3]　Yang ZH, Gong YZ, Xiao Q, Wang YW. DTS - A software defects testing system. In: Proc. of the 8th IEEE Int'l Working Conf. on Source Code Analysis and Manipulation. 2008. 269–270.

[4]　Kremenek T, Engler DR. Z-Ranking: using statistical analysis to counter the impact of static analysis approximations. In: Proc. of the 10th Int'l Static Analysis Symp.. 2003. 295–315.

[5]   Kim S, Ernst MD. Which warnings should I fix first? In: Proc. of the 6th Joint Meeting of the European Software Engineering Conf. and the ACM SIGSOFT Symp. on the Foundations of Software Engineering. New York: Association for Computing Machinery, 2007. 45–54.

[6]   Kang HJ, Aw KL, Lo D. Detecting false alarms from automatic static analysis tools: how far are we? In: Proc. of the 44th IEEE/ACM Int'l Conf. on Software Engineering. New York: Association for Computing Machinery, 2022. 698–709.

[7]   Kharkar A, Moghaddam RZ, Jin M, Liu XY, Shi X, Clement CB, Sundaresan N. Learning to reduce false positives in analytic bug detectors. In: Proc. of the 44th IEEE/ACM Int'l Conf. on Software Engineering. New York: Association for Computing Machinery, 2022. 1307–1316.

[8]   Johnson B, Song Y, Murphy-Hill ER, Bowdidge RW. Why don't software developers use static analysis tools to find bugs? In: Proc. of the 35th Int'l Conf. on Software Engineering. 2013. 672–681.

[9]   Sadowski C, Aftandilian E, Eagle A, Miller-Cushon L, Jaspan C. Lessons from building static analysis tools at Google. Commun. ACM, 2018,61(4):58–66.

[10]  Vassallo C, Panichella S, Palomba F, Proksch S, Gall HC, Zaidman A. How developers engage with static analysis tools in different contexts. Empir. Softw. Eng., 2020,25(2):1419–1457.

[11]  Li X, Zhou Y, Li MC, Chen YJ, Xu GQ, Wang LZ, Li XD. Automatically validating static memory leak warnings for C/C++ programs. Ruan Jian Xue Bao/Journal of Software, 2017,28(4):827–844 (in Chinese). http://www.jos.org.cn/1000-9825/5189.htm

[12]  Heckman SS, Williams LA. A systematic literature review of actionable alert identification techniques for automated static code analysis. Inf. Softw. Technol., 2011,53(4):363–387.

[13]  Yüksel U, Sözer H. Automated classification of static code analysis alerts: a case study. In: Proc. of the 2013 IEEE Int'l Conf. on Software Maintenance. 2013. 532–535.

[14]  Herzig K, Just S, Zeller A. It's not a bug, it's a feature: how misclassification impacts bug prediction. In: Proc. of the 35th Int'l Conf. on Software Engineering. 2013. 392–401.

[15]  Muske T, Serebrenik A. Survey of approaches for handling static analysis alarms. In: Proc. of the 16th IEEE Int'l Working Conf. on Source Code Analysis and Manipulation. 2016. 157–166.

[16]  Wang JJ, Wang S, Wang Q. Is there a "golden" feature set for static warning identification?: an experimental evaluation. In: Proc. of the 12th ACM/IEEE Int'l Symp. on Empirical Software Engineering and Measurement. New York: Association for Computing Machinery, 2018. 17:1–17:10.

[17]  Menzies T, Greenwald J, Frank A. Data mining static code attributes to learn defect predictors. IEEE Trans. on Software Eng., 2007,33(1):2–13.

[18]  Koc U, Wei SY, Foster JS, Carpuat M, Porter AA. An empirical assessment of machine learning approaches for triaging reports of a Java static analysis tool. In: Proc. of the 12th IEEE Int'l Conf. on Software Testing, Verification and Validation. 2019. 288–299.

[19]  Li H, Li XH, Chen X, Xie XF, Mu YZ, Feng ZY. Cross-project defect prediction via ASTToken2Vec and BLSTM-based neural network. In: Proc. of the 2019 Int'l Joint Conf. on Neural Networks. 2019. 1–8.

[20]  Zhao YY, Wang YW, Zhang YW, Zhang DL, Gong YZ, Jin DH. ST-TLF: Cross-version defect prediction framework based transfer learning. Inf. Softw. Technol., 2022,149:106939.

[21]  Jin Z, Liu F, Li G. Program comprehension: present and future. Ruan Jian Xue Bao/Journal of Software, 2019,30(1):110–126 (in Chinese). http://www.jos.org.cn/1000-9825/5643.htm

[22]  Mou LL, Li G, Zhang L, Wang T, Jin Z. Convolutional neural networks over tree structures for programming language processing. In: Proc. of the 30th AAAI Conf. on Artificial Intelligence. 2016. 1287–1293.

[23]  Wang S, Liu TY, Tan L. Automatically learning semantic features for defect prediction. In Proc. of the 38th Int'l Conf. on Software Engineering. New York: Association for Computing Machinery, 2016. 297–308.

[24]  Xing Y, Qian XM, Guan Y, Yang B, Zhang YW. Cross-project defect prediction based on G-LSTM model. Pattern Recognit. Lett., 2022,160:50–57.

[25]  White M, Vendome C, Vásquez ML, Poshyvanyk D. Toward deep learning software repositories. In: Proc. of the 12th IEEE/ACM Working Conf. on Mining Software Repositories. 2015. 334–345.

[26]  Fan G, Wu RX, Shi QK, Xiao X, Zhou JG, Zhang C. Smoke: scalable path-sensitive memory leak detection for millions of lines of code. In Proc. of the 41st Int'l Conf. on Software Engineering. 2019. 72–82.

[27] Le W, Soffa ML. Refining buffer overflow detection via demand-driven path-sensitive analysis. In Proc. of the 7th ACM SIGPLAN-SIGSOFT Workshop on Program Analysis for Software Tools and Engineering. New York: Association for Computing Machinery, 2007. 63–68.

[28] Ma XT, Yan JW, Yan J, Zhang J. Reorganizing and optimizing post-inspection on suspicious bug reports in path-sensitive analysis. In Proc. of the 19th IEEE Int'l Conf. on Software Quality, Reliability and Security. 2019. 260–271.

[29] Feng ZY, Guo DY, Tang DY, Duan N, Feng XC, Gong M, Shou LJ, Qin B, Liu T, Jiang DX, Zhou M. CodeBERT: a pre-trained model for programming and natural languages. In: Findings of the Association for Computational Linguistics: EMNLP. 2020. 1536–1547.

[30] Pan K, Kim S, Whitehead J. Bug classification using program slicing metrics. In: Proc. of the 6th IEEE Int'l Working Conf. on Source Code Analysis and Manipulation. 2006. 31–42.

[31] Ruthruff JR, Penix J, Morgenthaler D, Elbaum SG, Rothermel G. Predicting accurate and actionable static analysis warnings: an experimental approach. In Proc. of the 30th Int'l Conf. on Software Engineering. New York: Association for Computing Machinery, 2008. 341–350.

[32] Heckman SS, Williams LA. A model building process for identifying actionable static analysis alerts. In: Proc. of the 2nd Int'l Conf. on Software Testing Verification and Validation. 2009. 161–170.

[33] Hanam Q, Tan L, Holmes R, Lam P. Finding patterns in static analysis alerts: improving actionable alert ranking. In: Proc. of the 11th Working Conf. on Mining Software Repositories. 2014. 152–161.

[34] Yoon J, Jin M, Jung Y. Reducing false alarms from an industrial-strength static analyzer by SVM. In: Proc. of the 21st Asia-Pacific Software Engineering Conference. 2014. 3–6.

[35] Zhang YW, Xing Y, Gong YZ, Jin DH, Li HH, Liu F. A variable-level automated defect identification model based on machine learning. Soft Comput., 2020,24(2):1045–1061.

[36] Lee S, Hong S, Yi J, Kim T, Kim CJ, Yoo S. Classifying false positive static checker alarms in continuous integration using convolutional neural networks. In: Proc. of the 12th IEEE Int'l Conf. on Software Testing, Verification and Validation. 2019. 391–401.

[37] Xia WS, Li Y, Jia T, Wu ZH. BugIdentifier: an approach to identifying bugs via log mining for accelerating bug reporting stage. In Proc. of the 19th IEEE Int'l Conf. on Software Quality, Reliability and Security. 2019. 167–175.

[38] Li J, He PJ, Zhu JM, Lyu MR. Software defect prediction via convolutional neural network. In Proc. of the 2017 IEEE Int'l Conf. on Software Quality, Reliability and Security. 2017. 318–328.

[39] Zhang YW, Jin DH, Xing Y, Gong YZ. Automated defect identification via path analysis-based features with transfer learning. J. Syst. Softw., 2020,166:110585.

[40] Li ZQ, Jing XY, Zhu XK. Progress on approaches to software defect prediction. IET Softw., 2018,12(3):161–175.

[41] Herbold S, Trautsch A, Grabowski J. A comparative study to benchmark cross-project defect prediction approaches. IEEE Trans. Software Eng., 2018,44(9):811–833.

[42] Hosseini S, Turhan B, Gunarathna D. A systematic literature review and meta-analysis on cross project defect prediction. IEEE Trans. Software Eng., 2019,45(2):111–147.

[43] Zimmermann T, Nagappan N, Gall H, Giger E, Murphy B. Cross-project defect prediction: A large scale experiment on data vs. domain vs. process. In: Proc. of the 7th Joint Meeting of the European Software Engineering Conf. and the ACM SIGSOFT Symp. on the Foundations of Software Engineering. New York: Association for Computing Machinery, 2009. 91–100.

[44] Turhan B, Menzies T, Bener AB, Stefano J. On the relative value of cross-company and within-company data for defect prediction. Empir. Softw. Eng., 2009,14(5):540–578.

[45] Nam J, Pan SJ, Kim S. Transfer defect learning. In: Proc. of the 35th Int'l Conf. on Software Engineering. 2013. 382–391.

[46] Qiu SJ, Lu L, Cai ZY, Jiang SY. Cross-project defect prediction via transferable deep learning-generated and handcrafted features. In: Proc. of the 31st Int'l Conf. on Software Engineering and Knowledge Engineering. 2019. 431–552.

[47] Robertson S, Zaragoza H. The Probabilistic relevance framework: BM25 and beyond. Found. Trends Inf. Retr.,20093(4):333–389.

[48] Mikolov T, Sutskever I, Chen K, Corrado GS, Dean J. Distributed representations of words and phrases and their compositionality. In: Advances in Neural Information Processing Systems. 2013. 3111–3119.

[49] Devlin J, Chang MW, Lee K, Toutanova K. BERT: pre-training of deep bidirectional Transformers for language understanding. In: Proc. of the 2019 Conf. of the North American Chapter of the Association for Computational Linguistics: Human Language Technologies. 2019. 4171–4186.

[50] Vaswani A, Shazeer N, Parmar N, Uszkoreit J, Jones L, Gomez AN, Kaiser L, Polosukhin I. Attention is all you need. In: Advances in Neural Information Processing Systems. 2017. 5998–6008.

[51] Hoole AM, Traoré I, Delaitre A, Oliveira C. Improving vulnerability detection measurement: [test suites and software security assurance]. In: Proc. of the 20th Int'l Conf. on Evaluation and Assessment in Software Engineering. New York: Association for Computing Machinery, 2016. 27:1–27:10.

[52] Shiraishi S, Mohan V, Marimuthu H. Test suites for benchmarks of static analysis tools. In: Proc. of the 2015 IEEE Int'l Symp. on Software Reliability Engineering Workshops. 2015. 12–15.

[53] Boland T, Black PE. Juliet 1.1 C/C++ and Java test suite. Computer, 2012,45(10):88–90.

[54] Muske T, Talluri R, Serebrenik A. Repositioning of static analysis alarms. In: Proc. of the 27th ACM SIGSOFT Int'l Symp. on Software Testing and Analysis. New York: Association for Computing Machinery, 2018. 187–197.

[55] Zhang DL, Jin DH, Gong YZ, Wang Q, Dong YK, Zhang HL. Optimizing static analysis based on defect correlations. Ruan Jian Xue Bao/Journal of Software, 2014,25(2):386–399 (in Chinese). http://www.jos.org.cn/1000-9825/4538.htm

[56] Kocaguneli E, Tosun A, Bener AB, Turhan B, Caglayan B. Prest: an intelligent software metrics extraction, analysis and defect prediction tool. In: Proc. of the 21st Int'l Conf. on Software Engineering and Knowledge Engineering. 2009. 637–642.

[57] Bouckaert RR, Frank E, Hall MA, Holmes G, Pfahringer B, Reutemann P, Witten IH. Weka - experiences with a Java open-source project. J. Mach. Learn. Res., 2010,11:2533–2541.

[58] Arcuri A, Briand LC. A practical guide for using statistical tests to assess randomized algorithms in software engineering. In: Proc. of the 33rd Int'l Conf. on Software Engineering. 2011. 1–10.

[59] Tantithamthavorn C, McIntosh S, Hassan AE, Matsumoto K. The impact of automated parameter optimization on defect prediction models. IEEE Trans. Software Eng., 2019,45(7):683–711.

[60] Jureczko M, Madeyski L. Towards identifying software project clusters with regard to defect prediction. In: Proc. of the 6th Int'l Conf. On Predictive Models in Software Engineering. New York: Association for Computing Machinery, 2010. 9.

[61] Ghaffarian SM, Shahriari HR. Neural software vulnerability analysis using rich intermediate graph representations of programs. Inf. Sci., 2021,553:189–207.

**附中文参考文献:**

[11] 李筱,周严,李孟宸,陈园军,Xu Guo-Qiang,王林章,李宣东.C/C++程序静态内存泄漏警报自动确认方法.软件学报, 2017,28(4):827–844. http://www.jos.org.cn/1000-9825/5189.htm

[21] 金芝,刘芳,李戈.程序理解:现状与未来.软件学报,2019,30(1):110–126. http://www.jos.org.cn/1000-9825/5643.htm

[55] 张大林,金大海,宫云战,王前,董玉坤,张海龙.基于缺陷关联的静态分析优化.软件学报,2014,25(2):386–399. http://www.jos.org.cn/1000-9825/4538.htm